\documentclass[aps,pre,reprint]{revtex4-2}
\usepackage{amsmath}
\usepackage{graphicx}

\begin{document}


\title{Experiment and model for a Stokes layer in a strongly coupled dusty plasma}


\author{Jorge~Berumen}
\author{J.~Goree}
\affiliation{Department of Physics and Astronomy, University of Iowa, Iowa City, Iowa 52242, USA}


\date{\today}

\begin{abstract}	
A Stokes layer, which is a flow pattern that arises in a viscous fluid adjacent to an oscillatory boundary, was observed in an experiment using a two-dimensional strongly coupled dusty plasma. Liquid conditions were maintained using laser heating, while a separate laser manipulation applied an oscillatory shear that was localized and sinusoidal. The evolution of the resulting flow was analyzed using space-time diagrams. These figures provide an intuitive visualization of a Stokes layer, including features such as the depth of penetration and wavelength. Another feature, the characteristic speed for the penetration of the oscillatory flow, also appears prominently in space-time diagrams. To model the experiment, the Maxwell-fluid model of a Stokes layer was generalized to describe a two-phase liquid. In our experiment, the phases were gas and dust, where the dust cloud was viscoelastic due to strong Coulomb coupling. The model is found to agree with the experiment, in the appearance of the space-time diagrams and in the values of the characteristic speed, depth of penetration, and wavelength. 
\end{abstract}


\maketitle

\section{\label{secIntro}Introduction}

\begin{figure}
	\centering
	\includegraphics[width=1.0\columnwidth]{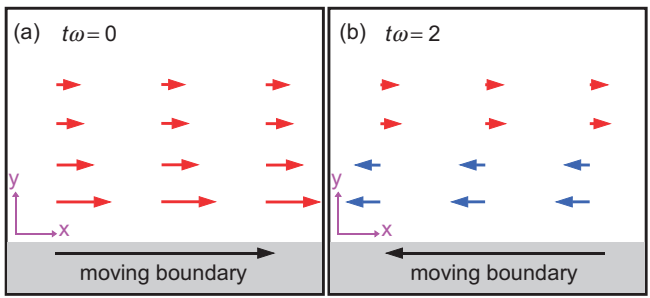}%
	\caption{\label{Fig1}Sketch of flow in a Stokes layer. An infinite planar boundary located at the bottom oscillates in the $\pm x$ direction. A flow develops in the adjoining fluid that fills the volume $y>0$. These sketches portray two different times, (a) $t\omega=0$ and (b) $t\omega=2$. Over time, as it oscillates, the flow reverses direction, as indicated by the different colors. This reversal occurs periodically with distance $y$ from the boundary, as is characterized by the parameter $\lambda$. The amplitude of the oscillation gradually weakens with increasing distance $y$, as characterized by the depth of penetration $\delta$.}
\end{figure}

A Stokes layer is a boundary layer that develops in a viscous fluid due to the oscillatory motion of an adjacent boundary plane~\cite{Schlichting2017}. This situation is also known as ``Stokes second problem'' and ``Stokes boundary layer.'' A flow pattern typical for a Stokes layer is sketched in Fig.~\ref{Fig1}.

\begin{figure}
	\centering
	\includegraphics[width=0.7\columnwidth]{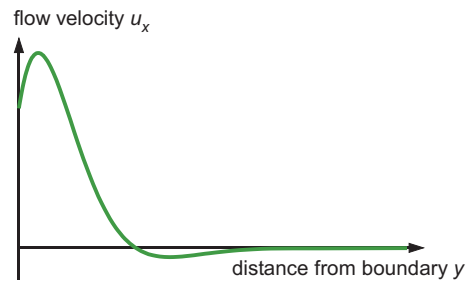}%
	\caption{\label{Fig2}Sketch of flow profile in a Stokes layer, at a particular time. In the Stokes layer literature, it is common to present the flow velocity $u_x$ in a one-dimensional graph, as sketched here.}
\end{figure}

In the literature, perhaps the most common graphical representation of a Stokes layer is a one-dimensional plot of flow velocity as a function of distance from the boundary~\cite{Amaratunga2020, Casanellas2012, Fetecau2008, Fetecau2009, Torralba2005}, at one moment in time, as sketched in Fig.~\ref{Fig2}. While this depiction of the flow profile displays the main spatial features of a Stokes layer, it is limited in its portrayal of the temporal evolution. 

A more complete visual characterization of a Stokes layer is provided by a space-time diagram. However, in our literature search, we found that space-time diagrams are uncommon, compared to one-dimensional graphs as in Fig.~\ref{Fig2}. Among the few previous publications using space-time diagrams that were revealed in our search, we found Ref.~\cite{Hack2015} for a purely viscous fluid and Ref.~\cite{Balmforth2009} for a viscoplastic fluid. 

In this paper, we expand on the previous literature for space-time diagrams. In Sec.~\ref{secST}, we show how these diagrams allow an intuitive visualization of the spatiotemporal evolution of the flow in a Stokes layer, and that it also allows an easy identification of the flow’s main features. 

A feature that we call a ``characteristic speed'' is particularly obvious in a space-time diagram. This speed, however, is not mentioned often in the literature. In fact, a speed is mentioned for a Stokes layer, as a characteristic measure of the flow profile, in only a few papers that we found in our literature search~\cite{Asghar2006, Hack2015, Hayat2004, Pritchard2011}. We did not expect this paucity of literature, when we commenced our search, because the existence of a characteristic speed is actually rather obvious by inspecting the theoretical solutions for the flow. Such a formula for the characteristic speed appears occasionally in the Stokes layer literature, for example in Ref.~\cite{Asghar2006} and Ref.~\cite{Hack2015}. In this paper, we devote Section~\ref{secSpd} to the characteristic speed, presenting its formula and demonstrating how it is revealed in a space-time diagram, for two kinds of fluids.

The most familiar textbook example of a Stokes layer is for a purely viscous fluid~\cite{Fetecau2008, Stokes1850, Torralba2005}; however, viscoelastic~\cite{Adler1949, Casanellas2011, Casanellas2012, Ferry1942, Fetecau2009, Khan2012, Mitran2008, Ortin2020, Schrag1977, Thurston1952, Thurston1959, Torralba2005, Vasquez2013} and non-Newtonian~\cite{Amaratunga2020, Balmer1980, Balmforth2009, Hayat2004, Khan2010, Rajagopal1982, Rajagopal1983} fluids have become more common in the research literature. In this paper we present an observation of a Stokes layer in another substance, a liquid-like dusty plasma, which has viscoelastic properties.

A chief result of this paper is that a dusty plasma can sustain a Stokes layer. We demonstrate this using an experiment. To provide both a spatial and temporal description of an oscillatory flow, we used video microscopy to analyze oscillatory flow profiles in a dusty plasma. These results allow us to present space-time diagrams and measure key parameters for the Stokes layer in our dusty plasma.

A dusty plasma is a mixture of four components: electrons, ions, neutral gas, and solid microscopic particles. These microscopic particles, which we call ``dust'' or ``dust particles,'' are typically a few microns in size. The relatively large size of these particles allows them to acquire large electric charges by collecting electrons and ions through collisions. In low-temperature laboratory plasmas, the charge is usually negative, with a magnitude on the order of $-10^{4}~e$.

The large charge of the dust particles offers several advantages. First, it allows the experimenter to levitate particles above a lower electrode, so that they are not in contact with any solid surface. Second, it provides a large interparticle repulsion. This repulsion leads to an interparticle potential energy that can be much greater than the kinetic energy. In other words, $\Gamma>1$ where $\Gamma$ is the Coulomb Coupling parameter,
\begin{equation}\label{Gamma}
\Gamma = \frac{Q^2}{4 \pi \epsilon_0 a_{\textrm{ws}} k_B T_k},
\end{equation}					
which is essentially the ratio of the interparticle potential energy and thermal kinetic energy. Here, $Q$ is the dust particle charge, $a_{\textrm{ws}}$ is the Wigner-Seitz radius, which characterizes the spacing between dust particles, and $T_k$ is the kinetic temperature of the dust particles. When $\Gamma>1$ the plasma is said to be strongly coupled, and the collection of charged dust particles does not behave like a rarefied gas, but instead like a liquid~\cite{Bonitz2014, Liu2007, Nosenko2006heat} or solid~\cite{Chu1994, Hayashi1994, Thomas1994, Totsuji2001}. In our experiment, we use a strongly coupled dusty plasma that has the properties of a liquid.

A liquid-like dusty plasma has qualities that are well suited for experiments. First, it is possible to track the motion of all the constituent particles, using video microscopy, as is done for example in some colloidal suspension experiments~\cite{Grier1994, Murray1989}. Second, the microscopic reorganizations of particles within this liquid are very slow, with time scales of order 0.1–1 s~\cite{Feng2012visco, Haralson2018, Hartman2011, Wong2018}, so that a video camera’s frame rate provides adequate temporal resolution. 

Flows of particles in a dusty plasma can be created by applying several kinds of forces. These include the Coulomb force~\cite{Heinrich2011, Jaiswal2015, Meyer2013, Meyer2014}, drag from a flowing gas~\cite{Arora2019, Carstensen2010, Hartman2013, Hartmann2019, Mitic2008}, gravity~\cite{Bailung2020}, and the radiation pressure force applied by a laser beam~\cite{Abbas2003, Liu2003}. The latter force is used in our experiment, as it was in previous experiments involving a shear flow. Some of those previous experiments used a steady shear flow~\cite{Chan2004, Chan2007, Feng2012temp, Fortov2012, Gavrikov2005, Haralson2016, Hartman2011, Io2009, Ivlev2007, Juan2001, Nosenko2004, Nosenko2012, Nosenko2013, Nosenko2020, Vaulina2007, Vorona2007} while in others the shear was suddenly switched on~\cite{Feng2010melt, LiuIEEE}. There have also been some experiments with a periodically modulated shear~\cite{Chan2007, Hartman2011}, but they were analyzed for purposes other than the identification of a Stokes layer. 

We also present models of a Stokes layer, under various conditions. We start in Sec.~\ref{secStokes} by reviewing well-known fluid descriptions for fluids that are either purely viscous or viscoelastic. We then extend the viscoelastic model to describe a two-phase fluid, which is descriptive of a dusty plasma. This extension, which takes into account frictional drag on a stationary gas background, results in the formulas presented in Sec.~\ref{sec2ph}. We compare our model to the experimental results in Sec.~\ref{secExpST}, showing that the model is reasonably accurate, in providing a qualitative description and in predicting quantitative parameters.

\section{\label{secStokes}Stokes Layer Theory}
\subsection{\label{subGeom}Geometry of the flow and boundary}

We consider a two-dimensional geometry, defined by the coordinates $(x,y)$. This geometry is applicable to physical systems that are truly two-dimensional (with all particles located on a single plane, as in our experiment) and it is also applicable to physical systems that fill a three-dimensional volume (as in experiments with a molecular fluid). The fluid's velocity is completely described by its $x$ component, $u_{x}(y,t)$, as the velocity in the $y$ direction is zero. 

The boundary is located at $y=0$, and can move only along the $x$ axis. This boundary is planar for a three-dimensional physical system, or linear for a truly two-dimensional system. To form a Stokes layer, this boundary is moved back and forth in the $\pm x$ directions. The movement of this boundary introduces an oscillatory shear that penetrates into the fluid. 

In our experiment, the fluid was a collection of dust particles that behaved like a liquid. Due to an electrical levitation, these dust particles filled only a single-layer two-dimensional cloud. They were not in contact with any solid surface, so that it was not practical to apply shear using a moving solid boundary. Instead, we applied shear by using laser beams to drive a localized flow of dust particles. The intensity of the laser beams oscillated, so that the shear had a sinusoidal variation with time. At the edges of our laser beams there was a localized back-and-forth flow of dust particles, analogous to the movement of fluid at the boundary of a solid surface in a traditional Stokes layer experiment using a molecular fluid. The shear laser setup is described in Sec.~\ref{subShear}.

\subsection{\label{subFlow}Flow profile in a planar Stokes layer}

In the textbook case of a Stokes layer, shear is externally applied by a moving boundary. Aside from the moving boundary, there are no other sources of motion, such as pressure gradients, in this simple textbook case. The boundary, which is denoted by a subscript b, has an oscillatory velocity along the $x$ axis that is specified as
\begin{equation}\label{eqBoundary}
	u_{b,x}(t)=\widetilde{U}\cos{\omega t},
\end{equation}
where $\omega$ is the frequency of oscillation. The amplitude $\widetilde{U}$ of this purely oscillatory motion is a constant. 

In contact with this boundary is a fluid, with a non-zero viscosity, filling a semi-infinite space. Within this fluid, momentum is transported, away from the moving boundary, by the diffusive effect of viscosity. A no-slip boundary condition is assumed, i.e., $u_x\left(y=0,t\right)=u_{b,x}(t)$. Within the fluid, the flow is not uniform, but instead has a profile with an amplitude that gradually decreases with distance from the boundary. For the textbook case of a Stokes layer, this oscillating flow is laminar.

The flow profile for this simple planar Stokes layer can be found by solving the Navier-Stokes equation~\cite{Stokes1850}. For the boundary condition of Eq.~(\ref{eqBoundary}), the flow profile is an exponentially suppressed sinusoidal oscillation,
\begin{equation}\label{eqSLbase}
	u_x\left(y,t\right)=\widetilde{U}e^{-y/\delta}\cos{\left(\omega t-\frac{2\pi}{\lambda}y\right)}.
\end{equation}

The Stokes layer solution of Eq.~(\ref{eqSLbase}) is applicable to several different kinds of fluids. For each of them, there is a different differential equation for the flow, but the solution has the form of Eq.~(\ref{eqSLbase}), as we will see in this paper.

The profile of the flow, as described by Eq.~(\ref{eqSLbase}), is sketched in Fig.~\ref{Fig1} for two different times. In this figure, we normalize time as $t\omega$, where $\omega$ is the externally imposed oscillation frequency for the boundary. 

We next explain the two parameters of a Stokes layer: $\delta$ and $\lambda$. 

The depth of penetration $\delta$ is the $e$-folding distance in Eq.~(\ref{eqSLbase}), which characterizes how deeply the oscillatory flow extends into the fluid from the moving boundary. Viscous transport is how the flow penetrates from the moving boundary into the liquid, but this transport is attenuated as the oscillatory momentum reaches greater distances into the fluid. Thus, viscosity serves two roles in a Stokes layer: it is required to transfer momentum from the boundary and into the fluid, and it ultimately diminishes this transfer of momentum at a characteristic distance $\delta$.

The parameter $\lambda$ is a measure of how, at a given moment in time, the flow direction reverses in the fluid. Since the boundary reverses directions periodically, and it takes time for momentum to be transported in the $y$ direction, the flow velocity at some distance from the boundary will also oscillate not only with time, but also with $y$. The value of $\lambda$ describes how this oscillation varies with distance from the boundary. The terminology used to describe $\lambda$ sometimes depends on the type of fluid. For a \textit{viscoelastic} fluid, where there is a restoring force, $\lambda$ is sometimes called a ``wavelength,''~\cite{Adler1949, Ferry1942, Ortin2020} and we will do the same since our dusty plasma has viscoelastic properties. On the other hand, in a \textit{purely viscous} fluid, it might be misleading to term $\lambda$ a wavelength since a wave requires a restoring force (to oppose inertia and thereby sustain the oscillation) and such a restoring force is absent in a purely viscous fluid. A graphical representation of $\lambda$ will be presented in Sec.~\ref{secST}.

\subsection{\label{subVis}Purely viscous fluid}

The dissipative effects in a purely viscous fluid are characterized by the shear viscosity $\eta_0$. One way of thinking of viscosity is that it governs a diffusion of momentum in a fluid. Due to this diffusion, the flow’s momentum is transferred in a direction perpendicular to the flow velocity. The momentum flux $\sigma_{\textrm{xy}}$, also called the shear stress, is driven by a velocity gradient $\partial u_x/\partial y$, as described by a constitutive relation $\sigma_{\textrm{xy}}=-\eta_0\partial u_x/\partial y$. The proportionality constant in this constitutive relation essentially defines the shear viscosity $\eta_0$.
 
The governing differential equation for a simple planar Stokes layer~\cite{Schlichting2017, Stokes1850} has the form of a diffusion equation:
\begin{equation}\label{eqDiffEqVis}
	\frac{\partial u_x}{\partial t}=\frac{\eta_0}{\rho}\frac{\partial^2u_x}{\partial y^2}.
\end{equation}
This equation is for a Newtonian liquid that is purely viscous (without elasticity) and has only a single phase (i.e., not a mixture of two different substances). Equation~(\ref{eqDiffEqVis}) can be derived from the Navier-Stokes equation for momentum, for a planar geometry, with no gradients in the pressure. The flow velocity $u_x$ does have gradients, which are normal to the boundary at $y=0$. The flow velocity $u_x(y,t)$ is specified at a distance $y$ from the boundary at a time $t$. 

When the boundary oscillates tangentially, as in the case of a Stokes layer, the solution of Eq.~(\ref{eqDiffEqVis}) has the form of Eq.~(\ref{eqSLbase}). In this solution for a semi-infinite fluid, there are two coefficients, the theoretical depth of penetration $\delta$ and wavelength $\lambda$. For a fluid that is purely viscous, they are~\cite{Fetecau2008, Schlichting2017, Stokes1850}
\begin{equation}\label{eqDvis}
	\delta_{\textrm{vis}}=\sqrt{\frac{2\eta_0}{\rho\omega}} ,
\end{equation}
and
\begin{equation}\label{eqLvis}
	\lambda_{\textrm{vis}}=2\pi\sqrt{\frac{2\eta_0}{\rho\omega}},
\end{equation}
where the subscript vis indicates a purely viscous fluid. 

\subsection{\label{subVE}Viscoelastic fluid}

Within a viscoelastic fluid, there are not only dissipative effects, but also energy-storing effects. To characterize this combination of properties, the viscosity is generalized as a frequency-dependent complex viscosity
\begin{equation}\label{eqFreqVisc}
	\eta\left(\omega\right)=\eta^{\prime}\left(\omega\right)-i\eta^{\prime\prime}(\omega),
\end{equation}
where the real term $\eta^\prime$ describes the viscous dissipation while the imaginary term $\eta^{\prime\prime}$ captures the energy-storing elastic effects. Viscoelasticity is inherently a time-dependent phenomenon, as reflected by the variable $\omega$. Physically, the time dependence arises because of an inherent time scale for particles to move microscopically amongst their neighbors. 

To describe the time dependence in a viscoelastic fluid, perhaps the simplest description is the Maxwell model, in which the inherent time scale is called the relaxation time $\tau$. For a Maxwell fluid, $\tau$ describes the temporal evolution of the relaxation of the stress $\sigma_{\textrm{xy}}$ after a sudden deformation. This temporal evolution is assumed to have a single exponential decay. This assumption, which was originally motivated by a mechanical arrangement of a dashpot in series with a spring, has been used successfully in describing many viscoelastic fluids~\cite{Buchanan2005, Cardinaux2002, Galvan-Miyoshi2008, Grimm2011, Sprakel2008, vanderGucht2003}, and strongly coupled dusty plasmas as well~\cite{Diaw2015, Donko2010, Feng2010visco, Feng2012visco, Goree2012, Hartman2011, Kaw1998}. 

Besides a frequency-dependent viscosity, as in Eq.~(\ref{eqFreqVisc}), a frequency-dependent shear modulus can also be used to describe a viscoelastic fluid. In a Maxwell fluid, these two descriptions are connected by the Maxwell relaxation time,
\begin{equation}\label{eqTau}
	\tau=\eta_0/G_{\infty}, 
\end{equation}
where $G_{\infty}$ is the high-frequency shear modulus and $\eta_0$ is the zero-frequency shear viscosity~\cite{March2002}. A substance with a great deal of elasticity will have a large value for $\tau$.	

Whether viscous or elastic effects dominate depends on the time scale for the evolution of the flow relative to the relaxation time $\tau$. To aid this comparison, we use a well-known dimensionless frequency, called the Deborah number
\begin{equation}\label{eqDe}
	De\equiv\omega\tau.
\end{equation}
We can say that viscous effects dominate at small $De$ (which can be attained at low frequencies)
\begin{equation}
\begin{aligned}
	\omega\tau & \ll 1\\
	De & \ll 1,
\end{aligned}
\end{equation}
while elastic effects are greatest at high $De$ (which can be attained at high frequencies)
\begin{equation}
\begin{aligned}
	\omega\tau & \gg 1 \\
	De & \gg 1.	
\end{aligned}
\end{equation}

The governing differential equation for a planar Stokes layer in a Maxwell fluid is~\cite{Ortin2020}
\begin{equation}\label{eqDiffEqVE}
	\left(1+\tau\frac{\partial}{\partial t}\right)\frac{\partial u_x}{\partial t}=\frac{\eta_0}{\rho}\frac{\partial^2u_x}{\partial y^2}.
\end{equation}
This generalization of Eq.~(\ref{eqDiffEqVis}) is intended for a viscoelastic fluid that has a single phase (i.e., not a mixture of two substances). The second term in the factor $\left(1+\tau\partial/\partial t\right)$ accounts for the memory-like effect of the elasticity of the fluid. In a viscoelastic fluid, microscopic memory is lost over time as particles rearrange themselves, and in a Maxwell fluid this memory loss is characterized by the time scale $\tau$. 

The solution of Eq.~(\ref{eqDiffEqVE}) for a viscoelastic fluid is again Eq.~(\ref{eqSLbase}), for the usual boundary condition with a tangential oscillation. However, the coefficients in this solution, the depth of penetration $\delta$ and wavelength $\lambda$, are different from those of a purely viscous fluid~\cite{Fetecau2009, Ortin2020}. 

The theoretical depth of penetration can be shown by solving Eq.~(\ref{eqDiffEqVE}), with the usual oscillating boundary condition of Eq.~(\ref{eqBoundary}), to be 
\begin{equation}\label{eqDve}
	\delta_{\textrm{ve}}=\sqrt{\frac{2\eta_0}{\rho\omega}}\left[\frac{1}{-De+\sqrt{1+De^2}}\right]^{1/2} . 
\end{equation}
This depth of penetration is greater than for a purely viscous liquid, Eq.~(\ref{eqDvis}). In other words, $\delta_{\textrm{ve}}>\delta_{\textrm{vis}}$, indicating that the oscillatory flow penetrates more deeply into a liquid with elastic properties than it does into a purely viscous liquid. 

The theoretical wavelength can be shown to be
\begin{equation}\label{eqLve}
	\lambda_{\textrm{ve}}=2\pi\sqrt{\frac{2\eta_0}{\rho\omega}}\left[\frac{1}{De+\sqrt{1+De^2}}\right]^{1/2} .	
\end{equation}
This wavelength is smaller than in Eq.~(\ref{eqLvis}) for a purely viscous fluid, due to the restoring effects of elasticity, i.e., $\lambda_{\textrm{ve}}<\lambda_{\textrm{vis}}$. In Eqs.~(\ref{eqDve}) and (\ref{eqLve}), the subscript ve specifies a viscoelastic fluid.

\section{\label{secST}Space-time Diagrams}

 \begin{figure*}
	\centering
	\includegraphics[width=2.0\columnwidth]{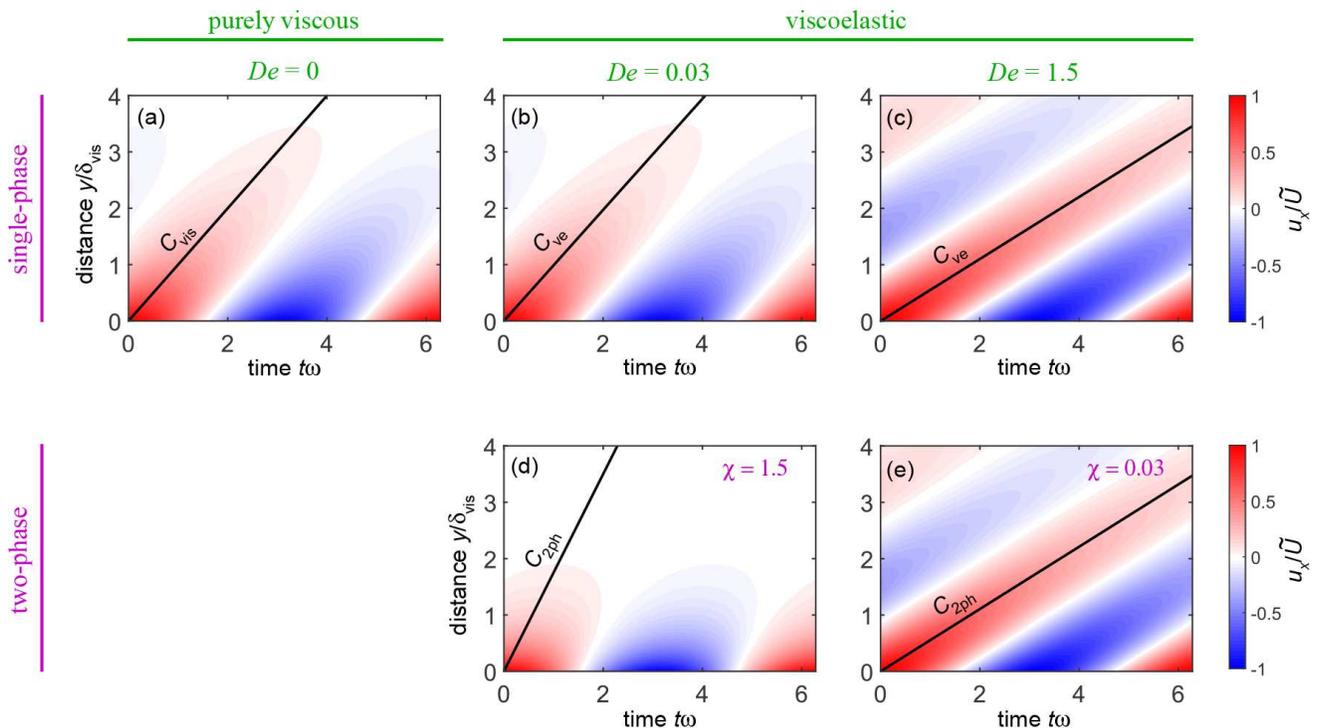}%
	\caption{\label{Fig3}(Color online) Space-time diagrams depicting the temporal evolution of flow velocity in a planar Stokes layer. The boundary, located at $y=0$, oscillates in the $\pm x$ direction at frequency $\omega$, as described by Eq.~(\ref{eqBoundary}). All five diagrams are plots of Eq.~(\ref{eqSLbase}), which is a solution of a differential equation, which is Eq.~(\ref{eqDiffEqVis}) for the \textit{purely viscous} fluid in (a), Eq.~(\ref{eqDiffEqVE}) for a \textit{viscoelastic} fluid in (b) and (c), and Eq.~(\ref{eqDiffEq2ph}) in (d) and (d) for a \textit{two-phase viscoelastic} fluid. The fading color, far from the boundary at $y=0$, shows how the flow decays with an $e$-folding distance $\delta$. Along the $y$ axis, a reversal of velocity reveals the wavelength $\lambda$. The dark line has a slope representing the characteristic speed $C$, as explained in Sec.~\ref{secSpd}. Time axes are normalized by the oscillation frequency $\omega$, while $y$ axes are normalized by the theoretical depth of penetration $\delta_{\textrm{vis}}$ for a purely viscous fluid, Eq.~(\ref{eqDvis}). Physical values for the experiment in Sec.~\ref{subExpparam} were used for these solutions: $\rho=\left(1.5\pm0.1\right)\times{10}^{-6}~\textrm{kg/m}^2$, $\eta_0=(3.5\pm0.4)\times{10}^{-12}~\textrm{kg~s}^{-1}$, $\tau=0.05$~s, and $\nu_g=0.97~\textrm{s}^{-1}$, with $\omega/2\pi=0.1$~Hz in (a), (b), (d); but $\omega/2\pi=5$~Hz in (c) and (e).}
\end{figure*}

A space-time diagram is a two-dimensional plot of the flow velocity profile, showing its spatiotemporal development. One axis is for a spatial coordinate $y$, and the other axis is for time. 

Space-time diagrams, as we will show in this section, are useful for characterizing the evolution of a Stokes layer's flow profile. Besides providing an intuitive way of visualizing a Stokes layer, these diagrams also allow an easy identification of features such as the depth of penetration and the wavelength. We are motivated to use space-time diagrams because they are familiar in the dusty plasma literature, not for Stokes layers, but for other purposes~\cite{Donko2020, Feng2010melt, Heinrich2011, LiuIEEE, Pikalev, Schwabe2011, Schwabe2014, Schwabe2020, Sheridan2008, Tsai2016, Williams2016, Zhdanov2015}, including the characterization of density waves~\cite{Bajaj2021, Himpel2012, Land2013, Melzer2020, Sarkar2016, Schwabe2008}. Some of these authors have used the terms periodograms or periodgrams for such plots.

Before we present space-time diagrams based on our experimental data in Sec.~\ref{secExpST}, we start by illustrating the concept of the diagram and its uses. As examples, in Fig.~\ref{Fig3} we present plots of Eq.~(\ref{eqSLbase}) for a planar Stokes layer. As in the sketches of Fig.~\ref{Fig1}, color is used to represent the direction of the flow, which reverses sinusoidally due to the boundary condition, Eq.~(\ref{eqBoundary}). In the top row of Fig.~\ref{Fig3}, space-time diagrams are shown for a purely viscous fluid ($De=0$), and for viscoelastic fluids. To illustrate the effects of viscoelasticity we consider two oscillation frequencies, corresponding to $De=0.03$ and $De=1.5$, for weak and strong elastic effects, respectively. 
 
The oscillations seen in these space-time diagrams have a periodicity in both time and space. Along the \textit{time} axis, the oscillation occurs at frequency $\omega$, which is externally provided at the boundary, and therefore persists indefinitely. 

Along the \textit{spatial} axis, an oscillation corresponds to $\lambda$. This spatial oscillation is most easily seen when elastic effects are strong, i.e., for high Deborah numbers as in Fig.~\ref{Fig3}(c). Besides the spatial \textit{oscillation}, there is also a spatial \textit{decay} with increasing distance from the boundary, and this is also easily seen in Fig.~\ref{Fig3}, as indicated by a color that fades with increasing $y$. This decay occurs over an $e$-folding distance $\delta$. The two parameters $\lambda$ and $\delta$ describe how the flow develops naturally, in response to the oscillating boundary and the properties of the fluid itself. 

We will use unsubscripted symbols $\lambda$ and $\delta$ in a general sense, for example to indicate values measured by empirical observation in an experiment. Theoretical values, on the other hand, are distinguished by subscripts: $\delta_{\textrm{vis}}$ and $\lambda_{\textrm{vis}}$ for a purely viscous single-phase fluid, and $\delta_{\textrm{ve}}$ and $\lambda_{\textrm{ve}}$ for a single-phase fluid that is viscoelastic. 

The half-wavelength $\lambda/2$ can be measured from a space-time diagram as the distance for the flow direction to reverse, at a given time. For example, at $t\omega=0$ in Fig.~\ref{Fig3}(a), there is a peak positive velocity at $y/\delta_{\textrm{vis}}=0$ as compared to a peak negative velocity at $y/\delta_{\textrm{vis}}=\pi$, as seen by the change in color. 

Elastic effects are expected to make the oscillations penetrate more deeply into a fluid, so that $\delta$ will be greater than for a purely viscous liquid. This prediction, which was quantified by Eq.~(\ref{eqDve}), can be seen in the space-time diagrams by comparing panels (a) and (c) in Fig.~\ref{Fig3}, for the limiting cases that are purely viscous and highly elastic, respectively. The oscillating flow's amplitude decays with distance from the boundary most slowly in the presence of high elasticity; this is seen in Fig.~\ref{Fig3}(c) where the oscillations are still easily distinguished at the maximum distance shown, $y/\delta_{\textrm{vis}}=4$. For comparison, the flow in the purely viscous liquid, Fig.~\ref{Fig3}(a), has almost completely decayed by that same distance.

Elastic effects also cause the oscillations to have a shorter wavelength $\lambda$. As we mentioned when discussing Eq.~(\ref{eqLve}), this trend is expected because elasticity provides a restoring force. This effect of elasticity on the wavelength can be rather profound, as seen in the space-time diagrams by comparing Fig.~\ref{Fig3}(a) and (c). 

\section{\label{secSpd}Characteristic Speed}

Another advantage of space-time diagrams is that they easily reveal a speed $C$ for the flow pattern as it penetrates the fluid. We call this a ``characteristic speed.'' As was noted by an earlier user of space-time diagrams~\cite{Hack2015}, this speed is identified as a tilted line, for a crest or trough of the oscillation, as we have drawn in Fig.~\ref{Fig3}. 

The existence of a characteristic speed is obvious from the cosine term of Eq.~(\ref{eqSLbase}). The cosine's argument indicates that for this planar Stokes layer, no matter what type of fluid, the characteristic speed is simply 
\begin{equation}\label{eqSpeed}
	C=\lambda\omega/2\pi. 
\end{equation}
Here, $\omega=2\pi f$ is the frequency of the externally applied motion that is imposed at the boundary, while $\lambda$ describes how these oscillations develop naturally as they move into the fluid.
 
The theoretical values for the characteristic speed are obtained by substituting $\lambda$ from Eq.~(\ref{eqLvis}) or (\ref{eqLve}) into Eq.~(\ref{eqSpeed}). For a purely viscous fluid, this yields
\begin{eqnarray}\label{eqSpeedVis}
	C_{\textrm{vis}}=\sqrt{\frac{2\eta_0\omega}{\rho}}, 
\end{eqnarray}
while for a viscoelastic Maxwell fluid it is 
\begin{equation}\label{eqSpeedVE}
	C_{\textrm{ve}}=\sqrt{\frac{2\eta_0\omega}{\rho}}\left[\frac{1}{De+\sqrt{1+De^2}}\right]^{1/2}.
\end{equation}
These two theoretical speeds are drawn as lines in the upper panels of Fig.~\ref{Fig3}. Along each line, the phase of the oscillations remains constant.

It might be misleading to call $C_{\textrm{vis}}$ a ``wave'' speed, just as it might be misleading to call $\lambda$ a ``wavelength,'' as we explained in Sec.~\ref{subFlow}. In a purely viscous fluid, the oscillations are not waves, because there is no restoring force, even if the oscillations do move through the medium at a definitive velocity. Only in the elastic limit is it compelling to describe the oscillation as a wave in the classical sense, where there is a restoring force that can oppose inertia.

In the elastic limit $De\gg 1$, the speed $C_{\textrm{ve}}$ in Eq.~(\ref{eqSpeedVE}) approaches a value $C_{\textrm{el}}$. Examining Eq.~(\ref{eqSpeedVE}), we see that $C_{\textrm{el}}=[\eta_0/\tau\rho]^{1/2}$, which can be combined with Eq.~(\ref{eqTau}) to yield
\begin{equation}\label{eqSpeedEl}
	C_{\textrm{el}}=\sqrt{\frac{G_{\infty}}{\rho}}.
\end{equation}
The expression for the elastic limit in Eq.~(\ref{eqSpeedEl}) has the same form as for the transverse sound speed in a solid, which would have a shear modulus $G$. (We note that a \textit{transverse} wave propagates freely in a \textit{solid}, and over a wide range of frequencies, but it propagates less readily in a \textit{liquid}, except at high frequencies~\cite{Hansen1986, Kalman2004, Nosenko2004}.)

In general, the characteristic speed $C$ is much less in the elastic limit $De\gg 1$ than in the viscous limit $De=0$. This tendency can be seen by comparing the top row of panels in Fig.~\ref{Fig3}, where the slope is smallest for the mostly elastic case of Fig.~\ref{Fig3}(c). No matter whether the medium is elastic or purely viscous, the oscillations are launched by the transverse oscillation of a boundary, but the resulting oscillations in the fluid have a character that is different, for the different kinds of fluid. In a purely viscous fluid, the oscillations result from momentum that is transported in the $y$ direction by \textit{diffusion}. In an elastic medium the momentum is carried by an \textit{inertia}, which is opposed by a restoring elastic force. It may not be intuitively obvious that adding elasticity to a fluid will cause its characteristic speed to be slower rather than faster, but that is indeed the case.

\section{\label{sec2ph}EXTENDING THE THEORY TO INCLUDE FRICTION ON A BACKGROUND MEDIUM}

\subsection{\label{subBackground}Background medium}

We next extend the well-known model for a Stokes layer in a viscoelastic fluid by taking into account an additional effect: frictional drag on another phase. In other words, we develop the model so that it is generally useful for a two-phase fluid. We do this because in a dusty plasma the dust particles occupy the same volume as a gas. This gas background, which is electrically neutral, can exert a frictional drag on the flow of dust particles. 

We will treat the case where the second phase is stationary. In the case of a dusty plasma, this means that the gas does not move, so that only the dust cloud flows. This is suitable for describing our experiment, where there was a tremendous difference in the areas of the large chamber walls as compared with the tiny surfaces of the few dust particles. Due to this disparity in surface areas, the gas makes much greater contact with the non-moving walls than with the moving dust particles, so that a flow of dust particles is not capable of pushing the gas into an overall movement along with the dust.

\subsection{\label{subDiffeq}Differential equation, depth of penetration and wavelength}

Extending the viscoelastic model for a Stokes layer by including a drag force term $\rho\nu_g u_x$ in Eq.~(\ref{eqDiffEqVE}), the governing equation for the planar viscoelastic Stokes layer becomes
\begin{equation}\label{eqDiffEq2ph}
	\left(1+\tau\frac{\partial}{\partial t}\right)\left[\frac{\partial u_x}{\partial t}+\nu_gu_x\right]=\frac{\eta_0}{\rho}\frac{\partial^2u_x}{\partial y^2}.
\end{equation}
Here, the term $\nu_gu_x$ accounts for the friction on the stationary gas background. The coefficient $\nu_g$ is a gas damping rate, defined as the dust particle's drag force divided by its momentum.

For the gas damping rate $\nu_g$, we use the Epstein theory~\cite{Epstein1924}. This well-known description of the frictional force is applicable to a sphere with a diameter much smaller than the mean-free-path of gas-gas collisions. In this model, the damping rate can be written as~\cite{Kananovich2020}
\begin{equation}\label{eqDampRate}
	\nu_g=d\frac{p_g}{\rho_dr_d}\sqrt{\frac{8m_g}{\pi k_BT_g}},
\end{equation}
where, $r_d$ and $\rho_d$ are the radius and mass density of the dust particles, while $p_g$, $m_g$, and $T_g$ are the pressure, atomic mass, and temperature of the background gas. The factor $d$ can take a value in the range 1.0 to 1.442, depending on the type of reflection that gas atoms experience on the surface of the dust particles~\cite{Epstein1924}. For our particles we use an experimentally obtained~\cite{Liu2003} value $d=1.26$.

Having added friction to the problem, the flow now has not just two, but three fundamental time scales. Besides the Maxwell relaxation time $\tau$ and the frequency $\omega$ of the boundary motion, there is also the gas damping rate $\nu_g$. We can compare these quantities using two dimensionless ratios: the Deborah number $De=\omega\tau$ along with another ratio that we introduce here,
\begin{equation}\label{eqChi}
	\chi\equiv\frac{\nu_g}{\omega}.
\end{equation}
Comparing gas friction and elasticity, in their effects on the flow profile, we note that each of these mechanisms has its greatest effect at opposite frequency limits. Elasticity, represented by the Deborah number $De=\omega\tau$, dominates at high $\omega$, as we discussed in Sec.~\ref{subVE}. Gas friction, on the other hand, has its greatest effect at low $\omega$, i.e., at large $\chi$, as we will show in Sec.~\ref{subST}.

The flow profile, i.e., the solution to Eq.~(\ref{eqDiffEq2ph}), is again Eq.~(\ref{eqSLbase}). What is different, compared to the single-phase fluids modeled in Sec.~\ref{secStokes}, are the effects of gas friction in the coefficients $\delta$ and $\lambda$, which we discuss next. 

The theoretical depth of penetration, for a viscoelastic Maxwell fluid with a friction on the stationary gas, can be shown by solving Eq.~(\ref{eqDiffEq2ph}) to be
\begin{equation}\label{eqD2ph}
	\delta_{\textrm{2ph}}=\sqrt{\frac{2\eta_0}{\rho\omega}}\left[\frac{1}{\left(\chi-De\right)+\sqrt{\left(\chi-De\right)^2+\left(1+\chi De\right)^2}}\right]^{1/2}.
\end{equation}

We can compare this expression to $\delta_{\textrm{ve}}$ in Eq.~(\ref{eqDve}), which also includes viscoelasticity but not gas friction. We make this comparison using the ratio
\begin{equation}\label{eqDcomp}
	\frac{\delta_{\textrm{2ph}}}{\delta_{\textrm{ve}}}=\left[\frac{-De+\sqrt{1+De^2}}{\left(\chi-De\right)+\sqrt{\left(\chi-De\right)^2+\left(1+\chi De\right)^2}}\right]^{1/2}.
\end{equation}
This ratio is less than unity, no matter what the frequency. In other words, friction on the background gas hinders the oscillatory Stokes layer from penetrating deeply into the fluid. 

The theoretical wavelength can be shown, by solving Eq.~(\ref{eqDiffEq2ph}), to be
\begin{equation}\label{eqL2ph}
	\lambda_{\textrm{2ph}}=2\pi\sqrt{\frac{2\eta_0}{\rho\omega}}\left[\frac{1}{-\left(\chi-De\right)+\sqrt{\left(\chi-De\right)^2+\left(1+\chi De\right)^2}}\right]^{1/2},
\end{equation}
for a viscoelastic Maxwell fluid with a friction on the stationary gas.
 
To assess the effect of friction, we compare Eq.~(\ref{eqL2ph}) to Eq.~(\ref{eqLve}), where the latter does not include friction, but both are intended for viscoelastic fluids. Again, presenting the comparison as a ratio, we write
\begin{equation}\label{eqLcomp}
	\frac{\lambda_{\textrm{2ph}}}{\lambda_{\textrm{ve}}}=\left[\frac{De+\sqrt{1+De^2}}{-\left(\chi-De\right)+\sqrt{\left(\chi-De\right)^2+\left(1+\chi De\right)^2}}\right]^{1/2}.
\end{equation}
We find that this ratio is greater than unity for all frequencies (except at infinite frequencies, where the ratio converges to unity). In physical terms, adding the effects of gas friction to a viscoelastic fluid tends to enhance the wavelength.

\subsection{\label{subST}Space-time diagram}

For a Stokes layer, instead of relying solely on equations and ratios, a space-time diagram offers a more graphical and intuitive way of identifying the effects of frictional drag. For this purpose, we again examine Fig.~\ref{Fig3}. To assess the role of gas friction in a viscoelastic fluid, we can compare the lower panels (d) and (e), which include friction, to the upper panels (b) and (c), which are for the frictionless case of a single-phase liquid.

At high frequency, i.e., high values of $De$, adding friction in (e) has no significant effect on the development and penetration of the flow profiles, as compared to the no-friction case of (c). As we mentioned above, in our discussion of the depth of penetration and wavelength, the effects of friction become comparatively weak at high $\omega$. Elasticity plays such a strong role at high frequency that other effects, such as friction, can play only a minor role. 

At low frequency, on the other hand, the effects of gas friction are more apparent. While the oscillatory flow profile is noticeable over a considerable distance $y$ in the absence of friction, Fig.~\ref{Fig3}(b), the oscillations penetrate a lesser distance when there is friction, Fig.~\ref{Fig3}(d). In other words, for a viscoelastic fluid, the depth of penetration $\delta$ is reduced by the presence of friction, as we discussed above regarding Eq.~(\ref{eqDcomp}).

\subsection{\label{subspeed}Characteristic speed}

To determine the characteristic speed in a two-phase viscoelastic Maxwell fluid, we can combine Eqs.~(\ref{eqSpeed}) and (\ref{eqL2ph}). This yields
\begin{equation}\label{eqSpeed2ph}
	C_{\textrm{2ph}}=\sqrt{\frac{2\eta_0\omega}{\rho}}\left[\frac{1}{-\left(\chi-De\right)+\sqrt{\left(\chi-De\right)^2+\left(1+\chi De\right)^2}}\right]^{1/2}.
\end{equation}

We can compare to a viscoelastic fluid without gas friction, Eq.~(\ref{eqSpeedVE}), by calculating a ratio
\begin{equation}\label{eqSpeedComp}
	\frac{C_{\textrm{2ph}}}{C_{\textrm{ve}}}=\left[\frac{De+\sqrt{1+De^2}}{-\left(\chi-De\right)+\sqrt{\left(\chi-De\right)^2+\left(1+\chi De\right)^2}}\right]^{1/2}.
\end{equation}
We note that this ratio is greater than unity for all frequencies (except at infinite frequencies where the ratio approaches unity). In other words, including gas friction tends to increase the characteristic speed in a viscoelastic fluid. This increase is most profound at low frequencies.

This increase in characteristic speed can be seen in our space-time diagrams. The slope is greater for the line in Fig.~\ref{Fig3}(d), taking friction into account, than it is in Fig.~\ref{Fig3}(b) without friction. These lines are plots of Eqs.~(\ref{eqSpeed2ph}) and (\ref{eqSpeedVE}) respectively. 

\subsection{\label{subBound}Boundary with an oscillation combined with constant velocity}

Until now, we have considered the most common boundary condition: a planar boundary at $y=0$ with a purely oscillatory motion and no added constant velocity, as described in Eq.~(\ref{eqBoundary}). 

In an experiment such as ours, there is a superposition of two motions at the boundary: an oscillatory motion $\widetilde{U}\cos{\omega t}$ and a constant velocity $U_0$. In this situation, we write the boundary condition at $y=0$ as
\begin{equation}\label{eqBoundary2}
	u_{b,x}(t)=U_0+\widetilde{U}\cos{\omega t}.
\end{equation}

The constant velocity actually poses no problem, in our analysis. We can show that the total flow profile $u_x\left(y,t\right)$, taking into account the constant velocity $U_0$, has almost the same form as Eq.~(\ref{eqSLbase}), which was for an oscillatory boundary with $U_0 = 0$. To show this, we start by separating the flow profile into a time-averaged component $U_x\left(y\right)=\langle u_x(y,t)\rangle _t$ and a fluctuating component ${\widetilde{u}}_x\left(y,t\right)=u_x(y,t)-U_x\left(y\right)$, i.e.,
\begin{equation}\label{eqTrial}
	u_x\left(y,t\right)=U_x\left(y\right)+{\widetilde{u}}_x(y,t). 
\end{equation}

It is shown in the Supplemental Material~\cite{SM} that the time-averaged component is
\begin{equation}\label{eqSteady}
	U_x\left(y\right)=U_0e^{-\sqrt{\rho\nu_g/\eta_0}y},
\end{equation}
and that the total flow profile is
\begin{equation}\label{eqTotal}
\begin{split}
	u_x\left(y,t\right) & =U_0e^{-\sqrt{\rho\nu_g/\eta_0}y} \\
	& +\widetilde{U}e^{-y/\delta_{\textrm{2ph}}}\cos{\left(\omega t-\frac{2\pi}{\lambda_{\textrm{2ph}}}y\right)}.
\end{split} 
\end{equation}

Comparing Eq.~(\ref{eqTotal}) to Eq.~(\ref{eqSLbase}), they have almost the same form. The only difference is in the first term of Eq.~(\ref{eqTotal}), which has no time dependence. Thus, space-time diagrams for the time-dependent portion of the experimentally measured flow velocity can be compared directly to our two-phase fluid model presented above.

\section{\label{secExp}Experiment}

 \begin{figure}
	\centering
	\includegraphics[width=1.0\columnwidth]{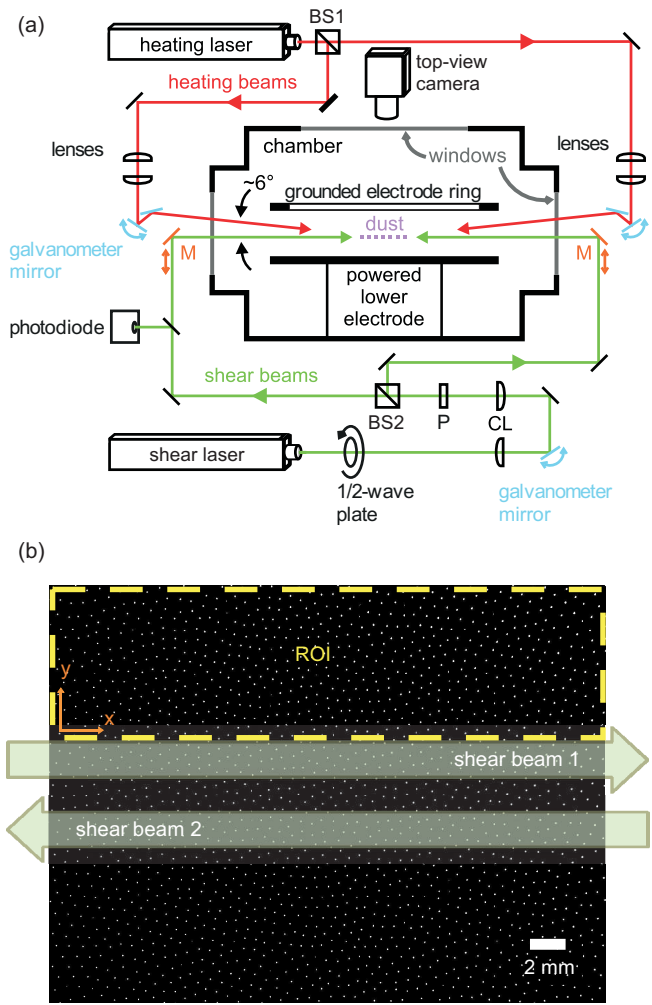}%
	\caption{\label{Fig4}(a) Sketch of the optical setup used to heat and apply shear to the dust cloud. The two laser beams in the lower portion of this diagram provided the shear. Before the shear beam was divided by a beam splitter (BS2), it was shaped by a combination of cylindrical lenses (CL) and a galvanometer mirror. It was also modulated sinusoidally in time, using a rotating half-wave plate and a linear polarizer (P), to allow a study of a Stokes layer rather than a steady-shear layer. The two heating beams in the upper portion were rastered over the entire dust-particle cloud. The top-view camera was our main diagnostic. (b) A single still image from the top-view camera, showing its field of view. The placement and width of the two shear beams are drawn to scale. The region of interest (ROI) was the portion of the image for $y>0$; this is where we measured particle velocities to yield the flow profile.}
\end{figure}

\subsection{\label{subApp}Apparatus}

We performed an experiment in a partially ionized Argon plasma. Using a modified GEC chamber, described in Ref.~\cite{Feng2011GK}, the gas pressure was regulated at 6~mTorr using a feedback controller and a capacitance manometer. Radio-frequency power at 13.56~MHz, with a peak-to-peak voltage of 122~V, was capacitively coupled to a lower electrode. The other electrode, which was grounded, consisted of the chamber walls as well as a smaller upper ring electrode. Levitation of the dust particles was aided by a DC self-bias of -61~V that developed on the lower electrode.

Dust particles formed a single-layer cloud after being introduced into the plasma. For our dust particles, we used melamine-formaldehyde microspheres, with a diameter of $2r_d=8.83~\mu\textrm{m}$. We calculated a particle's mass as $m_d=5.45\times{10}^{-13}$~kg, using the mass density $\rho_d=1.51~\textrm{g/cm}^3$ specified by the manufacturer. (In this calculation, we neglect mass loss due to exposure to vacuum and plasma conditions, which has been reported to be about 10\% of the original mass~\cite{Carstensen2011, Mikikian2003, Pavlu2004}.) We introduced about 5700~microspheres. They became levitated 9~mm above the lower electrode, and filled a circular area of diameter 64~mm. We verified using a side-view camera that there was only a single layer, with no particles above or below. 

As our main diagnostic, the dust particles were imaged from above. To illuminate the entire dust-particle cloud, light from a 577-nm laser was shaped into a horizontal sheet by a cylindrical lens. The top-view camera was a 12-bit Phantom Miro M120. Its 105-mm focal-length lens was fitted with an interference filter that admitted scattered laser light at 577~nm while blocking other wavelengths. This top-view camera was positioned above a window, at the top of the chamber. Its field of view was $24\times32$~mm containing about 2100~particles in the dust cloud's center. 

The laser heating the setup is sketched in Fig.~\ref{Fig4}(a). A pair of 532-nm laser beams was generated by a laser, operated at 12~W power and divided by a beamsplitter cube. Each beam was rastered in an arc-shaped pattern~\cite{Haralson2016IEEE}, so that impulses of momentum were imparted intermittently to dust particles. The arc-shaped pattern traced over a rectangular region including the entire dust-particle cloud. Using this laser heating, we were able to melt the crystal and sustain steady liquid conditions. Further details of the heating method and setup are provided in in Ref.~\cite{Haralson2016}.

To verify that we achieved liquid conditions, we compared the Coulomb coupling parameter $\Gamma$ to the melting point. The theoretical melting point~\cite{Hartmann2005} is $\Gamma_m=153$, for $\kappa=0.7$. For our experiment, $\Gamma$ was $78\pm6$, as calculated using Eq.~(\ref{Gamma}) along with the experimentally obtained parameters in Sec.~\ref{subExpparam}. Since $\Gamma\approx0.5\Gamma_m$, our temperature was about twice the melting point, so that the conditions were liquid.

\subsection{\label{subShear}Localized application of shear}

As we mentioned in Sec.~\ref{subGeom}, the oscillatory boundary in our Stokes layer experiment was not an external solid surface that oscillated in position. Instead we applied an oscillatory shear internally, using a pair of laser beams pointing oppositely in the $\pm x$ direction, with a gap between them, as shown in Fig.~\ref{Fig4}(b). A flow was produced, within each laser beam, due to the radiation pressure force. The laser beams were modulated only in intensity, but not in direction, driving a flow at $y=0$, i.e., at the edge of one laser beam. At that location, the velocity varied as in Eq.~(\ref{eqBoundary2}), with $\lvert U_0\rvert\geq\lvert\widetilde{U}\rvert$ and the same sign for both $U_0$ and $\widetilde{U}$. 

For applying shear, the laser manipulation setup was a modification of the one used in Ref.~\cite{Haralson2016}. Our shear laser was separate from the heating laser, although both were operated at the same wavelength of 532~nm, and both were split into two beams. The setup is shown in Fig.~\ref{Fig4}(a). The shear beams were shaped as horizontal ribbons, separated by a gap of 2~mm. These two laser beams were directed oppositely to serve two purposes: avoiding an overall horizontal displacement of the entire cloud, and forming a uniform shear region between the two beams. The width and height of the two ribbon-like beams were 2~mm and 1~mm, respectively.

Our simultaneous use of two kinds of laser manipulation, for heating and shear, has advantages as compared to applying shear alone~\cite{Haralson2016}. While it is possible to both melt a crystal and drive a flow using shear alone, there are two unwanted effects in the resulting liquid: shear thinning and spatial inhomogeneity. Those effects are avoided in our setup by applying a weaker shear. Our use of a weaker shear was made possible by melting the crystal separately, using the simultaneous application of our laser heating.

Two different regions were analyzed, when we applied the shear. Most importantly, for our Stokes layer analysis in Sec.~\ref{secExpST}, the region of interest (ROI) was the region \textit{outside} the gap between the shear beams, as in Fig.~\ref{Fig4}(b). The flow within this ROI is the result of a viscous momentum transfer from the oscillatory flow within Shear Beam 1. The high-velocity flow at $y=0$ serves as our analog to the moving boundary in a traditional Stokes layer. Additionally, for a separate analysis in Sec.~\ref{subExpparam} to obtain the shear viscosity $\eta_0$, the region analyzed was the gap between the two shear beams, where the shear was uniform.

To modulate the power of the shear laser, so that it varied sinusoidally with time, we used a pair of optical components: a rotating half-wave plate followed by a linear polarizer, which was stationary. These two components were positioned between the laser and the beamsplitter cube, as shown in Fig.~\ref{Fig4}(a) and in the Supplemental Material~\cite{SM}. With this arrangement, both shear beams had identical modulations. To precisely control the frequency of this modulation, we used a stepper motor to rotate the half-wave plate at a frequency $f/4$. We confirmed that the resulting modulation was sinusoidal, with a frequency $f$. It was necessary also to observe the phase of the modulation, for synchronization purposes, and this was accomplished using a photodiode detector. When the photodiode's output crossed a threshold value, it triggered a pulse.

We synchronized the top-view camera with the modulation of the shear laser. In particular, we controlled the camera to match both the frequency and phase of the laser modulation. To match the frequency, we used a master oscillator that controlled two pulse generators: one for triggering each camera frame, and the other for the stepper motor. To match the phase, we started the camera's recording when there was a photodiode trigger. A diagram of the camera triggering setup is provided in the Supplemental Material~\cite{SM}.

\subsection{\label{subProc}Procedure}

Experimental runs were carried out for two purposes: obtaining the space-time diagrams and measuring parameters of our dust cloud. We will describe these runs next, explaining how they required using the instrumentation differently.

The runs for obtaining our space-time diagrams used all of the features in our laser manipulation setup. Laser heating provided liquid conditions. The shear laser beams were used with sinusoidal modulation. The modulation frequency $f$ was chosen as 0.5~Hz, 1~Hz, or 2~Hz. (Such a low frequency is practical due to the slow time scale for microscopic reorganizations of dust particles in our liquid.) For each frequency, we performed ten runs. Due to our use of synchronization, as explained above, the ten runs were carried out under repeatable conditions, so that we could average the time series of their flow profiles.

The runs for measuring parameters were done three ways. First, our ``crystal runs'' were performed without laser heating and without shear. These crystal runs were used to obtain the areal number density by counting particles in the camera's full field of view. We then used the areal number density to obtain the Wigner-Seitz radius $a_{\textrm{ws}}$ and the areal mass density $\rho$. The crystal runs also provided the data required for our phonon-spectrum analysis~\cite{Nunomura2002phonon}, which we used to obtain the particle charge $Q$, dust plasma frequency $\omega_{\textrm{pd}}$, and shielding parameter $\kappa$. All of these parameters are expected to have the same value in a liquid as in a crystal. Second, our ``no-shear runs'' were performed with laser heating, but without shear, to obtain the average kinetic temperature $T_k$ of the dust particles. We used this temperature to calculate the value of the Coulomb coupling parameter $\Gamma$, which allowed us to confirm that liquid conditions were attained, as we explained above. Our no-shear runs also allowed us to obtain the Maxwell relaxation time $\tau$ of our liquid, as we explain later. Third, in our ``steady-shear runs'', we used both laser heating and shear, but the shear was maintained at a steady level by not rotating the half-wave plate. These steady-shear runs yielded a measurement of the shear viscosity $\eta_0$, as explained later.

Four repetitions of the runs for measuring parameters were performed during the course of the experiment. These repetitions allowed us to average a parameter's value over four observations and yielded an estimate of its uncertainty. We spaced the four repetitions widely in time, at the experiment's beginning and seven~hours later at its end, as well as twice in between. This scheme allowed us to confirm that there were no overall trends in the experimental conditions.

The top-view camera was operated at 100~frames per second for runs in liquid conditions, and 50~frames per second for the less demanding crystal conditions. These frame rates were chosen to meet two requirements: minimizing errors and allowing phase-resolved measurements. The latter required a camera frame rate that was an integer multiple of the laser modulation frequency $f$. To minimize velocity errors, we followed the prescription of Feng \textit{et al.}~\cite{Feng2011PTV}, which is intended to minimize the combination of acceleration error and random error. 

\subsection{\label{subExpdata}Obtaining flow profiles from particle data}

Our data analysis will center on a continuum description of the flow of dust particles. We obtain our flow data by analyzing video images as we describe here.

We started with a bit-map image corresponding to a single video frame, as in Fig.~\ref{Fig4}(b). We analyzed each frame to obtain the position ${(x}_i,y_i)$ of each particle $i$. This position measurement was done with sub-pixel accuracy by using the moment method, optimized as recommended in Ref.~\cite{Feng2007}. We mention here three of these optimization steps. First, we adjusted the camera lens so that each particle filled many pixels, with minimal aberration or distortion. Second, we reduced background noise in images by subtracting dark-field images. The latter were recorded under the same illumination conditions but without particles. Third, the threshold level in the moment method was selected to minimize errors both from pixel-locking and from random variations in pixel brightness. The result of these steps, which were carried out using ImageJ software~\cite{ImageJ}, were the positions of particles.

Particle-tracking velocimetry was used to obtain the velocity ${(v}_{i,x},v_{i,y})$ of each particle $i$. The algorithm we used was simply subtracting a particle's position in two consecutive frames and dividing by the time interval between frames. This standard method requires identifying the same particle in two consecutive frames, which was possible because we used a sufficiently high camera frame rate. 

The resulting particle-level data, with positions and velocities of individual particles, was used for several purposes. The positions yielded parameters for the number density, while the dispersion of velocities yielded the kinetic temperature. In combination, the positions and velocities were used to obtain the phonon spectrum and relaxation time, as described below. Besides those purposes, we also used the particle velocity data to obtain the flow velocity.

The flow velocity $u_x$ is a continuum description, which we obtained from the particle data. In essence, we converted our experimental data from the particle paradigm to the continuum paradigm. The key step in this conversion was binning. The region of interest ROI, in each video frame, was split into bins, which were rectangles that were thin in the $y$ direction. Since $x$ is an ignorable coordinate, and we wished to average over that direction, the bins were extended in the $x$ direction across the full width of the ROI. The flow velocity within each bin was obtained by averaging the velocities of particles located within, using a weighting algorithm. Repeating this process for each frame yielded our experimental flow profile, $u_x(y,t)$. We chose the bin width as $\Delta x=0.17$~mm to yield a sufficient resolution of about $a_{\textrm{ws}}/2$, while allowing adequate counting statistics in each bin.

The weighting algorithm that we used, in assigning particle velocities to a bin, was cloud-in-cell interpolation~\cite{Birdsall1991}. In this method, each particle contributes not just to one bin, but to the nearest two bins. The weight assigned to each bin is proportional to the distance to the center of that bin. The advantage of this algorithm is that it reduces noise resulting from a particle moving across the boundary dividing the two bins. Instead of an abrupt change in the obtained value of $u_x(y,t)$, there is only a gradual change as a single particle moves across a boundary between bins. The inputs for this interpolation are $y_i$ and $v_{i,x}$ for a particle.

\subsection{\label{subExpparam}Parameters of the dust particle cloud}

We required experimentally measured values of dusty plasma parameters to characterize our dust-particle cloud, as well as to use our two-phase fluid model. We report these parameters for three kinds of runs: crystal, no-shear, and steady-shear runs. 

From our crystal runs, we obtained the values of the following parameters. The areal number density of the dust-particle cloud was $2.7\pm{0.1~\textrm{mm}}^{-2}$. The Wigner-Seitz radius, which characterizes the interparticle spacing, was $a_{\textrm{ws}}=0.34\pm0.01$~mm. The areal mass density of the cloud was $\rho=\left(1.5\pm0.1\right)\times{10}^{-6}~\textrm{kg/{m}}^2$. Using the phonon-spectrum method~\cite{Nunomura2002disp, Nunomura2002phonon, Wang2001}, we found the dust plasma frequency was $\omega_{\textrm{pd}}=\left[Q^2/2\pi\epsilon_0m_da_{\textrm{ws}}^3\right]^{1/2}=81.0\pm0.4~\textrm{{s}}^{-1}$ and the shielding parameter was $\kappa=a_{\textrm{ws}}/\lambda_{\textrm{D}}=0.70\pm0.02$, which yields the screening length $\lambda_{\textrm{D}}=0.49\pm0.01$~mm. The dust particle charge, obtained from $\omega_{\textrm{pd}}$, was $Q=(17\ 500\pm500)~e$. The error values for all these parameters were calculated from the dispersion of values obtained in various runs. 

Our no-shear runs were used to measure the kinetic temperature $T_k$ of the dust-particle cloud, which was required to calculate the Coulomb coupling parameter $\Gamma$ for liquid conditions. We calculated the kinetic temperature from the dispersion of the particle velocities~\cite{Schablinski}, using $T_k=\left(m_d/2k_B\right)\langle\left(\textbf{v}_i\left(\textbf{x}_i,t\right)-\bar{\textbf{v}}\left(t\right)\right)^2\rangle_{i,t}$, where $\bar{\textbf{v}}\left(t\right)=\langle\textbf{v}_i\left(\textbf{x}_i,t\right)\rangle_{i,t}$. The subscripts of the angle brackets indicate that the averaging was performed over particles $i$ and over time $t$. (This temperature is a kinetic temperature associated with the random movement of dust particles; it is not descriptive of the polymer substance within a particle, which was much colder.) For our experiment, under liquid conditions, the kinetic temperature was $T_k=\left(2.0\pm0.1\right)\times{10}^5~\textrm{K}$. The corresponding value of the Coulomb coupling parameter $\Gamma$, calculated from Eq.~(\ref{Gamma}), was $\Gamma=78\pm6$. 

Analyzing our no-shear runs, we also estimated the Maxwell relaxation time $\tau$. We did this using the same steps as in Ref.~\cite{Donko2010, Feng2012visco}. First, a time series of the shear stress $\sigma_{\textrm{xy}}(t)$ was computed from particle-level data for positions and velocities, as described in Ref.~\cite{Feng2012visco}. Second, the autocorrelation function of the shear stress was calculated as $A_\eta(t)=\left\langle\sigma_{\textrm{xy}}\left(t\right)\sigma_{\textrm{xy}}\left(0\right)\right\rangle_t$. Third, the frequency-dependent viscosity $\eta(\omega)$ was obtained by calculating the Laplace-Fourier transform of $A_\eta(t)$ and using the generalized Green-Kubo relation to yield $\eta\left(\omega\right)$. Finally, $\tau$ was estimated by fitting the Maxwell model's analytic expression for $\eta(\omega)$ to the experimental result, where $\tau$ was a free parameter, as described in Ref.~\cite{Feng2012visco}. This method requires a liquid under steady conditions, which we were able to achieve with our heating setup. For our experiment, the resulting value of the relaxation time is $\tau=0.05$~s. 

We used steady-shear runs to determine the shear viscosity $\eta_0$, which is a required input in our two-phase fluid model. The steps for obtaining $\eta_0$ from particle-level data include calculations of both the shear stress $\sigma_{\textrm{xy}}$, and the shear rate $\gamma$, as described in Ref.~\cite{Haralson2016}. These two values were obtained by analyzing the shear region between the two shear beams. The shear stress was obtained using Eq.~(8) of Ref.~\cite{Haralson2016}, with an input of particle positions and velocities, along with values of $\omega_{\textrm{pd}}$ and $\lambda_{\textrm{D}}$. The shear rate was calculated as the spatial derivative of the flow velocity within that shear region, where the shear was nearly uniform. We then calculated the shear viscosity as the ratio $\eta_0={-\sigma}_{\textrm{xy}}/\gamma=\left(3.5\pm0.4\right)\times{10}^{-12}~\textrm{kg~{s}}^{-1}$.

We note that motion of particles within our cloud was underdamped. This situation is different from a colloidal suspension, which is overdamped due to filling the space between particles with a massive solvent. In our experiment, the space between particles was filled with rarefied gas, so that the particles experienced a frictional damping rate of only $\nu_g=0.97~\textrm{{s}}^{-1}$, as calculated using Eq.~(\ref{eqDampRate}). This value for the frictional damping rate is two orders of magnitude less than $\omega_{\textrm{pd}}$, meaning that friction on the gas background, while certainly a factor, was not a dominant factor in the particle motion. 

\subsection{\label{subExpflow}Preparing flow profiles for space-time diagrams}

Before producing the space-time diagrams, we prepared the experimental flow profiles. To do this, we first obtained the fluctuating component of the flow velocity ${\widetilde{u}}_x$, and we then performed a phase-resolved averaging to improve the signal-to-noise ratio. We next describe details of these two steps.

The fluctuating component of the flow ${\widetilde{u}}_x$ was obtained from the time series data for the flow velocity in each bin. Using Eq.~(\ref{eqTrial}), the experimental fluctuating component is obtained as ${\widetilde{u}}_x\left(y,t\right)=u_x\left(y,t\right)-U_x\left(y\right)$. The latter value is obtained by averaging the time series $u_x\left(y,t\right)$ separately within each spatial bin. This procedure was performed for each oscillatory shear run.

Phase-resolved averaging was then performed to improve the signal-to-noise ratio for the experimental ${\widetilde{u}}_x\left(y,t\right)$. For each run, we split the time series ${\widetilde{u}}_x\left(y,t\right)$ into non-overlapping time segments, each with a duration of one modulation cycle. Each segment had the same phase because the frames recorded by the camera were synchronized with the modulation, and there was an integer number of frames in each cycle. Thus, in each segment the first value corresponded to the same phase of the sinusoidal modulation. Likewise, the second value in each segment corresponded to another phase of the modulation, and so on for one complete cycle. To obtain a phase-resolved average of ${\widetilde{u}}_x\left(y,t\right)$, we averaged the first value in each segment over all ten runs, and then similarly averaged the second value in each segment, and so on. This phase-resolved flow profile is what we use as the input for preparing our space-time diagram.

\section{\label{secExpST}Experimental Space-time Diagram}

 \begin{figure*}
	\centering
	\includegraphics[width=2.0\columnwidth]{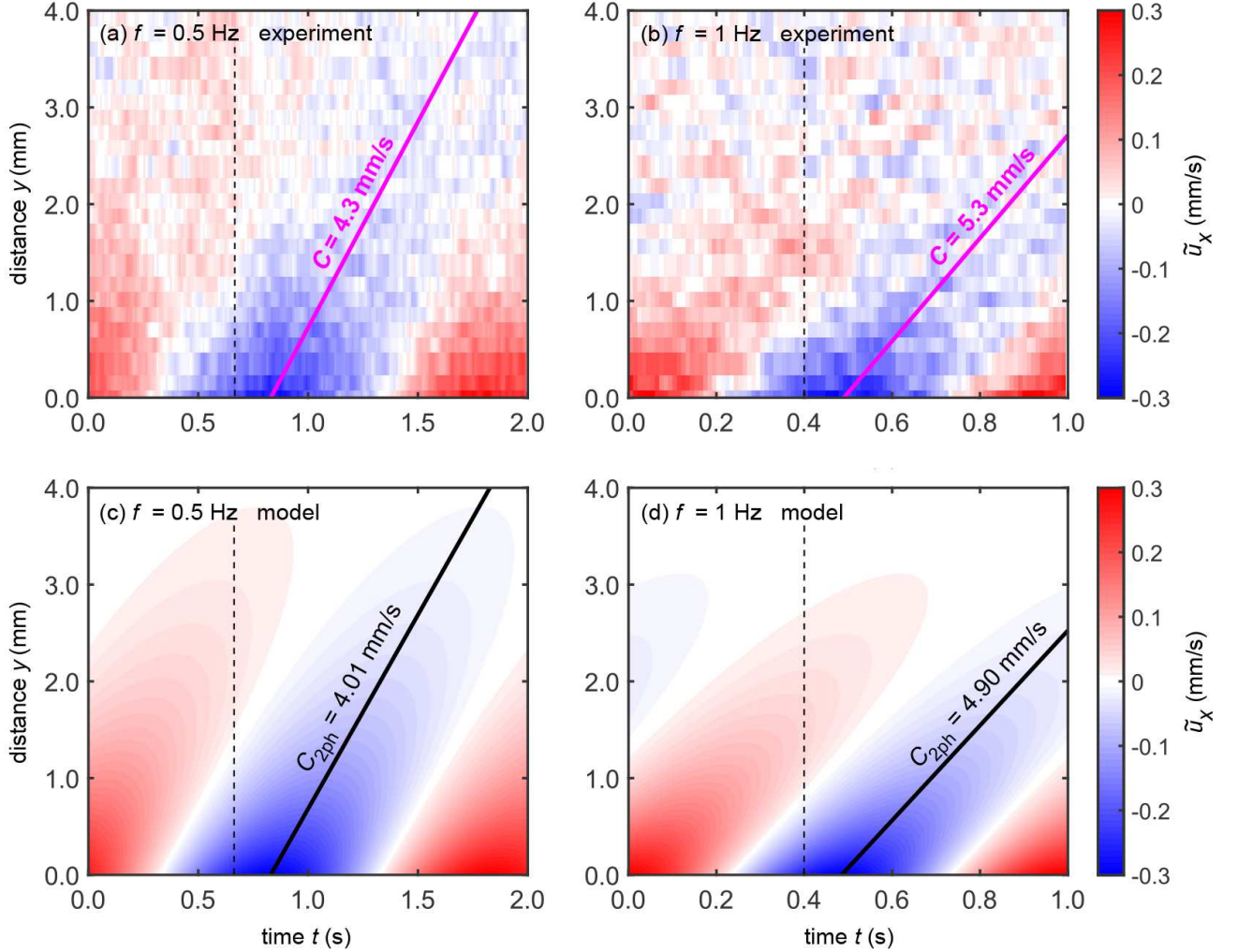}%
	\caption{\label{Fig5}(Color online) Space-time diagrams obtained from experimental data (top row) and our two-phase fluid model (bottom row), for modulation frequencies of 0.5~Hz (left column) and 1~Hz (right column). Zero time corresponds to a trigger event in the experiment. The color bar scale for ${\widetilde{u}}_x$ is the same for all panels. In the experimental space-time diagrams, the main features of a Stokes layer are apparent: the finite penetration of the flow can be seen as fading of the color with increasing distance; a reversal of the flow direction can be seen as a reversing color with distance from the boundary, for example along the vertical dashed line in each diagram; the characteristic speed is indicated by the slope of the solid lines. In addition to obtaining $C$, these experimental diagrams also allow obtaining values for $\delta$ and $\lambda$. The theoretical diagrams were obtained by combining Eqs.~(\ref{eqSLbase}), (\ref{eqD2ph}), and (\ref{eqL2ph}), while the theoretical characteristic speed $C_{\textrm{2ph}}$ was calculated using Eq.~(\ref{eqSpeed2ph}), all with an input of experimental parameters listed in the main text.}
\end{figure*}

For the experiment, our primary results are space-time diagrams of flow profiles in Fig.~\ref{Fig5}. These plots are our results for modulation frequencies of 0.5~Hz and 1~Hz. (The space-time diagram for 2~Hz, presented in the Supplemental Material~\cite{SM}, is noisier.) These plots illustrate the spatial and temporal evolution of the fluctuating component of the flow, ${\widetilde{u}}_x(y,t)$. They also allow us to obtain quantitative values for $C$, $\delta$, and $\lambda$, as well as qualitatively identifying three significant features, as we report next.

\subsection{\label{subExpAnal}Analysis}

The first feature of a Stokes layer that we identify is a spatial decay of the flow’s amplitude, as seen in Fig.~\ref{Fig5}(a) and (b). Qualitatively, this decay appears as a fading of colors, with increasing distance from the boundary at $y=0$. Quantitatively, the depth of penetration was obtained by fitting the experimental data to an exponential decay, yielding $\delta=1.00\pm0.05$~mm for $f=0.5$~Hz and $0.85\pm0.05$~mm for $f=1$~Hz. 

The second feature is a characteristic speed $C$. This speed is apparent in the experimental data as an overall tilt in the spatiotemporal flow pattern. For example, one can observe the development in time of the location of peak positive flow or peak negative flow. To quantify this characteristic speed, we performed a correlation analysis, yielding $C=4.3$~mm/s for 0.5~Hz and $5.3$~mm/s for 1~Hz, shown as the slope of the solid lines in Fig.~\ref{Fig5}(a) and (b). 

The third feature is the reversal of the fluctuating flow’s direction. This reversal can be identified qualitatively by examining data along a vertical line in Fig.~\ref{Fig5}(a) and (b). For example, along the vertical dashed line in each plot, the flow velocity ${\widetilde{u}}_x$ is negative at small distances in the lower quarter of the graph, but positive at larger distances. To quantify the wavelength, which captures the scale length for this flow reversal, we used Eq.~(\ref{eqSpeed}), yielding $\lambda=8.5$~mm for 0.5~Hz and $5.3$~mm for 1~Hz.

\subsection{\label{subExpComp}Comparison to two-phase fluid model}

Comparing these experimental results to the predictions of our two-phase fluid model, we find that they agree. For this comparison, in the model we use a combination of Eqs.~(\ref{eqSLbase}), (\ref{eqD2ph}), and (\ref{eqL2ph}) along with Eq.~(\ref{eqSpeed2ph}) for the characteristic speed, with an input of the experimentally measured values of $\omega$, $\eta_0$, $\rho$, $\tau$, and $\nu_g$ as reported in Sec.~\ref{subExpparam}.
 
Qualitatively, we see the same spatiotemporal development of the flow profiles in the experiment and the model, in Fig.~\ref{Fig5}. Both exhibit the same three features mentioned above: a spatial decay seen as a fading of colors, a skew that reflects the characteristic speed, and a reversal of color with distance from the boundary indicating a reversal in the flow’s direction.
 
Quantitatively, we find reasonable agreement for the key parameters for the Stokes layer. The depth of penetration predicted by the model is $\delta_{2\textrm{ph}}=1.10$~mm for 0.5~Hz and 0.90~mm for 1~Hz, differing from the experiment by 10\% and 6\%, respectively. The characteristic speed predicted by the model is $C_{2\textrm{ph}}=4.01$~mm/s for 0.5~Hz and 4.90~mm/s for 1~Hz, which compared to the experiment differed by 7\% and 8\%, respectively. Those same percentage differences apply also to the wavelengths, which in the model were $\lambda_{2\textrm{ph}}=8.01$~mm for 0.5~Hz and 4.90~mm for 1~Hz.

\section{\label{subConc}Conclusions}

A Stokes layer was observed in a dusty plasma experiment. The collection of particles in a two-dimensional cloud behaved as a liquid with a steady temperature, due to heating with laser manipulation. Using another kind of laser manipulation, which applied a sustained sinusoidally modulated shear within the dust cloud, we drove an oscillatory flow. This shear manipulation serves as an analog of a sinusoidally moving boundary in a traditional Stokes layer experiment. Using video microscopy, we measured particle velocities to obtain a spatiotemporal characterization of the flow pattern produced by the oscillatory shear manipulation. 

We used space-time diagrams to present the spatiotemporal flow profiles. One axis of these diagrams represents time, over a cycle of the sinusoidal modulation. The other axis is for distance from the boundary (or in the case of our experiment it is distance from the edge of the shear laser beam.) In these space-time diagrams, the features of the Stokes layer stand out. Beyond the overall flow pattern with its oscillatory reversal of flow, there are some specific features of a Stokes layer that are easily identified by inspecting a space-time diagram. These features include the depth of penetration (for the exponential decay of the flow pattern with distance) and a characteristic speed.

The characteristic speed $C$ is seen as an overall tilt or skew in the spatiotemporal profiles for the oscillatory flow pattern, in the space-time diagram. The crests and troughs of the oscillation follow a line with a distinctive slope in the space-time diagram, and that slope is the characteristic speed. While we find this speed to be a distinctive feature of a Stokes layer, it seems not to be mentioned often in the fluid-mechanics literature. 

Fluid models were presented to illustrate the use of space-time diagrams, and to compare to our experiment. In particular, we developed a two-phase fluid model, which includes frictional forces and viscoelastic effects. This model is intended to describe a liquid-like collection of particles in a two-dimensional dusty plasma, but it may also be applicable more generally to a two-phase fluid in other substances, in which a background phase (gas in our experiment) remains stationary while exerting a frictional force on the other phase (small solid particles in our experiment). Viscoelasticity is incorporated in this model using a complex viscosity that includes a relaxation time that captures the effect of elasticity; these elastic effects can be turned off simply by setting the relaxation time to zero, if desired to model a purely viscous phase.

Comparing the experiment and our two-phase fluid model, we find reasonable agreement for the Stokes layer. The overall spatiotemporal pattern is captured well by the model. Quantitatively, we find agreement within 10\% for the values of the depth of penetration, characteristic speed, and wavelength.

\begin{acknowledgments}
	This work was supported by U.S. Department of Energy grant DE-SG0014566, the Army Research Office under MURI Grant W911NF-18-1-0240, and NASA-JPL subcontracts 1573629 and 1663801. We would like to thank B. Liu and A. Kananovich for helpful discussions.	
\end{acknowledgments}


\begin{thebibliography}{49}
	
	\bibitem{Schlichting2017}
	H. Schlichting and K. Gersten, \textit{Boundary-Layer Theory}, 9th ed. (Springer-Verlag Berlin Heidelberg, 2017).
	
	\bibitem{Amaratunga2020}
	M. Amaratunga, H. A. Rabenjafimanantsoa, and R.~W.~Time, J. Non-Newton. Fluid Mech. {\bf 277}, 104236 (2020).
	
	\bibitem{Casanellas2012}
	L. Casanellas and J. Ort\'in, Rheol. Acta {\bf 51}, 545 (2012).
	
	\bibitem{Fetecau2008}
	C. Fetecau, D. Vieru, and C. Fetecau, Int. J. Non-Linear Mech. {\bf 43}, 451 (2008).
	
	\bibitem{Fetecau2009}
	C. Fetecau, M. Jamil, C.~Fetecau, and I.~Siddique, Int. J. Non-Linear Mech. {\bf 44}, 1085 (2009).
	
	\bibitem{Torralba2005}
	M. Torralba, J. R. Castrej\'on-Pita, A. A. Castrej\'on-Pita, G.~Huelsz, J.~A.~del~R\'io, and J.~Ort\'in, Phys. Rev. E {\bf 72}, 016308 (2005).
	
	\bibitem{Hack2015}
	J. P. Hack and T. A. Zaki, J. Fluid Mech. {\bf 778}, 389 (2015).
	
	\bibitem{Balmforth2009}
	N. J. Balmforth, Y. Forterre, and O. Pouliquen, J. Non-Newton. Fluid Mech. {\bf 158}, 46 (2009).
	
	\bibitem{Asghar2006}
	S. Asghar, S. Nadeem, K. Hanif, and T. Hayat, Math. Probl. Eng. {\bf 2006}, 1 (2006).
	
	\bibitem{Hayat2004}
	T. Hayat, S. Asghar, and A. M. Siddiqui, Appl. Math. Comput. {\bf 148}, 697 (2004).
	
	\bibitem{Pritchard2011}
	D. Pritchard, C. R. McArdle, and S. K. Wilson, J. Non-Newton. Fluid Mech. {\bf 166}, 745 (2011).	
	
	\bibitem{Stokes1850}
	G. G. Stokes, Trans. Cambridge Philos. Soc. {\bf IX}, 8 (1850).	
	
	\bibitem{Adler1949}
	F. T. Adler, W. M. Sawyer, and J. D. Ferry, J. Appl. Phys. {\bf 20}, 1036 (1949).
	
	\bibitem{Casanellas2011}
	L. Casanellas and J. Ort\'in, J. Non-Newton. Fluid Mech. {\bf 166}, 1315 (2011).	
	
	\bibitem{Ferry1942}
	J. D. Ferry, J. Am. Chem. Soc. {\bf 64}, 1323 (1942).
	
	\bibitem{Khan2012}
	M. Khan, M. Arshad, and A. Anjum, Nucl. Eng. Des. {\bf 243}, 20 (2012).
	
	\bibitem{Mitran2008}
	S. M. Mitran, M. G. Forest, L. Yao, B.~Lindley, and D.~B.~Hill, J. Non-Newton. Fluid Mech. {\bf 154}, 120 (2008).
	
	\bibitem{Ortin2020}
	J. Ort\'in, Phil. Trans. R. Soc. A {\bf 378}:20190521 (2020).
	
	\bibitem{Schrag1977}
	J. L. Schrag, Trans. Soc. Rheol. {\bf 21}, 399 (1977).
	
	\bibitem{Thurston1952}
	G. B. Thurston, J. Acoust. Soc. Am. {\bf 24}, 653 (1952).
	
	\bibitem{Thurston1959}
	G. B. Thurston, J. Appl. Phys. {\bf 30}, 1855 (1959).
	
	\bibitem{Vasquez2013}
	P. A. Vasquez, Y. Jin, K. Vuong, D. B. Hill, and M.~Gregory Forest, J. Non-Newton. Fluid Mech. {\bf 196}, 36 (2013).		
	
	\bibitem{Balmer1980}
	R. T. Balmer and M. A. Florina, J. Non-Newton. Fluid Mech. {\bf 7,} 189 (1980).	
	
	\bibitem{Khan2010}
	M. Khan, A. Anjum, C. Fetecau, and H. Qi, Math. Comput. Model. {\bf 51}, 682 (2010).
	
	\bibitem{Rajagopal1982}
	K. R. Rajagopal, Int. J. Non-Linear Mech. {\bf 17}, 369 (1982).
	
	\bibitem{Rajagopal1983}
	K. R. Rajagopal and T. Y. Na, Acta Mech. {\bf 48}, 233 (1983).		
	
	\bibitem{Bonitz2014}
	M. Bonitz, J. Lopez, K. Becker, and H. Thomsen, \textit{Complex Plasmas: Scientific Challenges and Technological Opportunities}, Vol. 82 (Springer, New York, 2014).
	
	\bibitem{Liu2007}
	B. Liu and J. Goree, Phys. Rev. E {\bf 75}, 016405 (2007).
	
	\bibitem{Nosenko2006heat}
	V. Nosenko, J. Goree, and A. Piel, Phys. Plasmas {\bf 13}, 032106 (2006).
	
	\bibitem{Chu1994}
	J. H. Chu and L.~I, Phys. Rev. Lett. {\bf 72}, 4009 (1994).
	
	\bibitem{Hayashi1994}
	Y. Hayashi and K. Tachibana, Jpn. J. Appl. Phys {\bf 33}, 804 (1994).
	
	\bibitem{Thomas1994}
	H. Thomas, G. E. Morfill, V. Demmel, J. Goree, B.~Feuerbacher, and D.~M\"ohlmann, Phys. Rev. Lett. {\bf 73}, 652 (1994).
	
	\bibitem{Totsuji2001}
	H. Totsuji, C. Totsuji, and K. Tsuruta, Phys. Rev. E {\bf 64}, 066402 (2001).
	
	\bibitem{Grier1994}
	D. G. Grier and C. A. Murray, J. Chem. Phys. {\bf 100}, 9088 (1994).
	
	\bibitem{Murray1989}
	C. A. Murray and R. A. Wenk, Phys. Rev. Lett. {\bf 62}, 1643 (1989).
	
	\bibitem{Feng2012visco}
	Y. Feng, J. Goree, and B. Liu, Phys. Rev. E {\bf 85}, 066402 (2012).
	
	\bibitem{Haralson2018}
	Z. Haralson, J. Goree, and R. Belousov, Phys. Rev. E {\bf 98}, 023201 (2018).
	
	\bibitem{Hartman2011}
	P. Hartmann, M. C. S\'andor, A. Kov\'acs, and Z. Donk\'o, Phys. Rev. E {\bf 84}, 016404 (2011).
	
	\bibitem{Wong2018}
	C. S. Wong, J. Goree, and Z. Haralson, IEEE Trans. Plasma Sci. {\bf 46}, 763 (2018).
	
	\bibitem{Heinrich2011}
	J. R. Heinrich, S. H. Kim, J. K. Meyer, and R. L. Merlino, Phys. Plasmas {\bf 18}, 113706 (2011).
	
	\bibitem{Jaiswal2015}
	S. Jaiswal, P. Bandyopadhyay, and A. Sen, Rev. Sci. Instrum. {\bf 86}, 113503 (2015).
	
	\bibitem{Meyer2013}
	J. K. Meyer, J. R. Heinrich, S. H. Kim, and R. L. Merlino, J. Plasma Phys. {\bf 79}, 677 (2013).
	
	\bibitem{Meyer2014}
	J. K. Meyer, R. L. Merlino, J. R. Heinrich, and S.~H.~Kim, IEEE Trans. Plasma Sci. {\bf 42}, 2690 (2014).
	
	\bibitem{Arora2019}
	G. Arora, P. Bandyopadhyay, M. G. Hariprasad, and A.~Sen, Phys. Plasmas {\bf 26}, 023701 (2019).
	
	\bibitem{Carstensen2010}
	J. Carstensen, F. Greiner, and A. Piel, Phys. Plasmas {\bf 17}, 083703 (2010).	
	
	\bibitem{Hartman2013}
	P. Hartmann, Z. Donk\'o, T. Ott, H. K\"ahlert, and M.~Bonitz, Phys. Rev. Lett. {\bf 111}, 155002 (2013).
	
	\bibitem{Hartmann2019}
	P. Hartmann, J. C. Reyes, E. G. Kostadinova, L.~S.~Matthews, T.~W.~Hyde, R.~U.~Masheyeva, K.~N.~Dzhumagulova, T.~S.~Ramazanov, T.~Ott, H.~K\"ahlert, M.~Bonitz, I.~Korolov, and Z.~Donk\'o, Phys. Rev. E {\bf 99}, 013203 (2019).
	
	\bibitem{Mitic2008}
	S. Mitic, R. S\"utterlin, A.~V.~Ivlev, H.~Höfner, M.~H.~Thoma, S.~Zhdanov, and G.~E.~Morfill, Phys. Rev. Lett. {\bf 101}, 235001 (2008).
	
	\bibitem{Bailung2020}
	Y. Bailung, B. Chutia, T. Deka, A. Boruah, S.~K.~Sharma, S.~Kumar, J.~Chutia, Y.~Nakamura, and H.~Bailung, Phys. Plasmas {\bf 27}, 123702 (2020).
	
	\bibitem{Abbas2003}
	M. M. Abbas \textit{et al.}, J. Geophys. Res. {\bf 108}, 1229 (2003).
	
	\bibitem{Liu2003}
	B. Liu, J. Goree, V. Nosenko, and L. Boufendi, Phys. Plasmas {\bf 10}, 9 (2003).
	
	\bibitem{Chan2004}
	C. L. Chan, W. Y. Woon, and L.~I, Phys. Rev. Lett. {\bf 93}, 220602 (2004).
	
	\bibitem{Chan2007}
	C. L. Chan and L.~I, Phys. Rev. Lett. {\bf 98}, 105002 (2007).
	
	\bibitem{Feng2012temp}
	Y. Feng, J. Goree, and B. Liu, Phys. Rev. Lett. {\bf 109}, 185002 (2012).
	
	\bibitem{Fortov2012}
	V. E. Fortov, O. F. Petrov, O. S. Vaulina, and R.~A.~Timirkhanov, Phys. Rev. Lett. {\bf 109}, 055002 (2012).
	
	\bibitem{Gavrikov2005}
	A. Gavrikov, I. Shakhova, A. Ivanov, O. Petrov, N.~Vorona, and V.~Fortov, Phys. Lett. Sect. A {\bf 336}, 378 (2005).
	
	\bibitem{Haralson2016}
	Z. Haralson and J. Goree, Phys. Plasmas {\bf 23}, 093703 (2016).
	
	\bibitem{Io2009}
	C. W. Io and L.~I, Phys. Rev. E {\bf 80}, 036401 (2009).
	
	\bibitem{Ivlev2007}
	A. V. Ivlev, V. Steinberg, R. Kompaneets, H. H\"ofner, I.~Sidorenko, and G.~E.~Morfill, Phys. Rev. Lett. {\bf 98}, 145003 (2007).
	
	\bibitem{Juan2001}
	W. T. Juan, M. H. Chen, and L.~I, Phys. Rev. E {\bf 64}, 016402 (2001).
	
	\bibitem{Nosenko2004}
	V. Nosenko and J. Goree, Phys. Rev. Lett. {\bf 93}, 155004 (2004).
	
	\bibitem{Nosenko2012}
	V. Nosenko, A. V. Ivlev, and G. E. Morfill, Phys. Rev. Lett. {\bf 108}, 135005 (2012).
	
	\bibitem{Nosenko2013}
	V. Nosenko, A. V. Ivlev, and G. E. Morfill, Phys. Rev. E {\bf 87}, 043115 (2013).
	
	\bibitem{Nosenko2020}
	V. Nosenko, M. Pustylnik, M. Rubin-Zuzic, A.~M.~Lipaev, A.~V.~Zobnin, A.~D.~Usachev, H.~M.~Thomas, M.~H.~Thoma, V.~E.~Fortov, O.~Kononenko, and A.~Ovchinin, Phys. Rev. Res. {\bf 2}, 033404 (2020).
	
	\bibitem{Vaulina2007}
	O. S. Vaulina, O. F. Petrov, A. V. Gavrikov, X.~G.~Adamovich, and V.~E.~Fortov, Phys. Lett. A {\bf 372}, 1096 (2007).
	
	\bibitem{Vorona2007}
	N. A. Vorona, A. V. Gavrikov, A. S. Ivanov, O.~F.~Petrov, V.~E.~Fortov, and I.~A.~Shakhova, J. Exp. Theor. Phys. {\bf 105}, 824 (2007).
	
	\bibitem{Feng2010melt}
	Y. Feng, J. Goree, and B. Liu, Phys. Rev. Lett. {\bf 104}, 165003 (2010).
	
	\bibitem{LiuIEEE}
	B. Liu, J. Goree, M. Y. Pustylnik, H. M. Thomas, V.~E.~Fortov, A.~M.~Lipaev, A.~D.~Usachev, F.~Petrov, A.~V~Zobnin, and M.~H.~Thoma, IEEE Trans. Plasma Sci. (submitted).
	
	\bibitem{Buchanan2005}
	M. Buchanan, M. Atakhorrami, J. F. Palierne, F.~C.~MacKintosh, and C.~F.~Schmidt, Phys. Rev. E {\bf 72}, 011504 (2005).
	
	\bibitem{Cardinaux2002}
	F. Cardinaux, L. Cipelletti, F. Scheffold, and P.~Schurtenberger, EPL {\bf 57}, 738 (2002).
	
	\bibitem{Galvan-Miyoshi2008}
	J. Galvan-Miyoshi, J. Delgado, and R. Castillo, Eur. Phys. J. E {\bf 26}, 369 (2008).
	
	\bibitem{Grimm2011}
	M. Grimm, S. Jeney, and T. Franosch, Soft Matter {\bf 7}, 2076 (2011).
	
	\bibitem{Sprakel2008}
	J. Sprakel, J. van der Gucht, M. A. Cohen Stuart, and N.~A.~M.~Besseling, Phys. Rev. E {\bf 77}, 061502 (2008).
	
	\bibitem{vanderGucht2003}
	J. van der Gucht, N. A. M. Besseling, W. Knoben, L.~Bouteiller, and M.~A.~Cohen~Stuart, Phys. Rev. E {\bf 67}, 051106 (2003).
	
	\bibitem{Diaw2015}
	A. Diaw and M. S. Murillo, Phys. Rev. E {\bf 92}, 013107 (2015).
	
	\bibitem{Donko2010}
	Z. Donk\'o, J. Goree, and P. Hartmann, Phys. Rev. E {\bf 81}, 056404 (2010).
	
	\bibitem{Feng2010visco}
	Y. Feng, J. Goree, and B. Liu, Phys. Rev. Lett. {\bf 105}, 025002 (2010).
	
	\bibitem{Goree2012}
	J. Goree, Z. Donk\'o, and P. Hartmann, Phys. Rev. E {\bf 85}, 066401 (2012).
	
	\bibitem{Kaw1998}
	P. K. Kaw and A. Sen, Phys. Plasmas {\bf 5}, 3552 (1998).
	
	\bibitem{March2002}
	N. H. March and M. P. Tosi, \textit{Introduction to Liquid State Physics}, (World Scientific, River Edge New Jersey, 2002), p. 286.
	
	\bibitem{Donko2020}
	Z. Donk\'o, P. Hartmann, R. U. Masheyeva, and K.~N.~Dzhumagulova, Contrib. to Plasma Phys. {\bf 60}:e201900197 (2020).
	
	\bibitem{Pikalev}
	A. Pikalev, M. Pustylnik, C.~R\"ath, and H.~Thomas, \textit{Heartbeat instability as auto-oscillation between dim and bright void regimes} (2021), arXiv:2103.06795 [physics.plasm-ph], URL \url{https://arxiv.org/abs/2103.06795}.
	
	\bibitem{Schwabe2011}
	M. Schwabe, S. Zhdanov, M. Rubin-Zuzic, A. Ivlev, H.~Thomas, and G.~Morfill, IEEE Trans. Plasma Sci. {\bf 39}, 2726 (2011).
	
	\bibitem{Schwabe2014}
	M. Schwabe, S. Zhdanov, C. R\"ath, D. B. Graves, H.~M.~Thomas, and G.~E.~Morfill, Phys. Rev. Lett. {\bf 112}, 115002 (2014).
	
	\bibitem{Schwabe2020}
	M. Schwabe {\it et al}., New J. Phys. {\bf 22}, 083079 (2020).
	
	\bibitem{Sheridan2008}
	T. E. Sheridan, V. Nosenko, and J. Goree, Phys. Plasmas {\bf 15}, 073703 (2008).
	
	\bibitem{Tsai2016}
	Y. Y. Tsai, J. Y. Tsai, and L.~I, Nat. Phys. {\bf 12}, 573 (2016).
	
	\bibitem{Williams2016}
	J. D. Williams, J. Plasma Phys. {\bf 82}, 615820302 (2016).
	
	\bibitem{Zhdanov2015}
	S. Zhdanov, M. Schwabe, C. R\"ath, H. M. Thomas, and G.~E.~Morfill, EPL {\bf 110}, 35001 (2015).
	
	\bibitem{Bajaj2021}
	P. Bajaj, S. Khrapak, V. Yaroshenko, M.~Schwabe, \textit{Spatial distribution of Dust Density Wave Properties in Fluid Complex Plasmas} (2021), arXiv:2105.12036 [physics.plasm-ph], URL \url{https://arxiv.org/abs/2105.12036}.
	
	\bibitem{Himpel2012}
	M. Himpel, C. Killer, B.~Buttensch\"on, and A.~Melzer, Phys. Plasmas {\bf 19}, 123704 (2012).
	
	\bibitem{Land2013}
	V. Land, A. Douglass, K. Qiao, Z.~Zhang, L.~S.~Matthews, and T.~Hyde, IEEE Trans. Plasma Sci. {\bf 41}, 799 (2013).
	
	\bibitem{Melzer2020}
	A. Melzer, H. Kr\"uger, S. Sch\"utt, and M.~Mulsow, Phys. Plasmas {\bf 27}, 033704 (2020).
	
	\bibitem{Sarkar2016}
	S. Sarkar, C. Barman, M. Mondal, M. Bose, and S.~Mukherjee, J. Phys. D. Appl. Phys. {\bf 49}, 205201 (2016).
	
	\bibitem{Schwabe2008}
	M. Schwabe {\it et al}., New J. Phys. {\bf 10}, 033037 (2008).
	
	\bibitem{Hansen1986}
	P. Hansen and I. R. McDonald, \textit{Theory of Simple Liquids}, 2nd ed. (Elsevier Academic, Amsterdam, 1986).
	
	\bibitem{Kalman2004}
	G. J. Kalman, P. Hartmann, Z. Donk\'o, and M. Rosenberg, Phys. Rev. Lett. {\bf 92}, 065001 (2004).
	
	\bibitem{Nosenko2006cutoff}
	V. Nosenko, J. Goree, and A. Piel, Phys. Rev. Lett. {\bf 97}, 115001 (2006).
	
	\bibitem{Epstein1924}
	P. S. Epstein, Phys. Rev. {\bf 23}, 710 (1924).
	
	\bibitem{Kananovich2020}
	A. Kananovich and J. Goree, Phys. Plasmas {\bf 27}, 113704 (2020).
	
	\bibitem{SM}
	See Supplemental Material at http:// Xxxxxxxxx
	
	\bibitem{Feng2011GK}
	Y. Feng, J. Goree, B. Liu, and E.~G.~D.~Cohen, Phys. Rev. E {\bf 84}, 046412 (2011).
	
	\bibitem{Carstensen2011}
	J. Carstensen, H. Jung, F. Greiner, and A. Piel, Phys. Plasmas {\bf 18}, 033701 (2011).
	
	\bibitem{Mikikian2003}
	M. Mikikian, L. Boufendi, A.~Bouchoule, H.~M.~Thomas, G.~E.~Morfill, A.~P.~Nefedov, and V.~E.~Fortov, New J. Phys. {\bf 5}, 19 (2003).
	
	\bibitem{Pavlu2004}
	J. Pavl\r{u}, A. Velyhan, I. Richterov\'a, Z.~N\v{e}me\v{c}ek, J.~\v{S}afr\'ankov\'a, I.~\v{C}erm\'ak, and P.~\v{Z}ilav\'y, IEEE Trans. Plasma Sci. {\bf 32}, 704 (2004).
	
	\bibitem{Haralson2016IEEE}
	Z. Haralson and J. Goree, IEEE Trans. Plasma Sci. {\bf 44}, 549 (2016).
	
	\bibitem{Hartmann2005}
	P. Hartmann, G. J. Kalman, Z.~Donk\'o, and K.~Kutasi, Phys. Rev. E {\bf 72}, 026409 (2005).
	
	\bibitem{Nunomura2002phonon}
	S. Nunomura, J. Goree, S. Hu, X. Wang, and A.~Bhattacharjee, Phys. Rev. E {\bf 65}, 066402 (2002).
	
	\bibitem{Feng2011PTV}
	Y. Feng, J. Goree, and B. Liu, Rev. Sci. Instrum. {\bf 82}, 053707 (2011).
	
	\bibitem{Feng2007}
	Y. Feng, J. Goree, and B. Liu, Rev. Sci. Instrum. {\bf 78}, 053704 (2007).
	
	\bibitem{ImageJ}
	W. S. Rasband, computer code ImageJ, U. S. National Institutes of Health, Bethesda, Maryland, USA, https://imagej.nih.gov/ij/, 1997-2018.
	
	\bibitem{Birdsall1991}
	C. K. Birdsall and A. B. Langdon, \textit{Plasma Physics via Computer Simulation} (IOP Publishing Ltd., New York, 1991).
	
	\bibitem{Nunomura2002disp}
	S. Nunomura, J. Goree, S. Hu, X. Wang, A.~Bhattacharjee, and K.~Avinash, Phys. Rev. Lett. {\bf 89}, 035001 (2002).
	
	\bibitem{Wang2001}
	X. Wang, A. Bhattacharjee, and S.~Hu, Phys. Rev. Lett. {\bf 86}, 2569 (2001).
	
	\bibitem{Schablinski}
	J. Schablinski, D. Block, A. Piel, A. Melzer, H.~Thomsen, H.~K\"ahlert, and M.~Bonitz, Phys. Plasmas {\bf 19}, 013705 (2012).
	
\end{thebibliography}
\end{document}